\documentclass{article}
\usepackage{csquotes}
\usepackage{amsmath}
\usepackage{multicol}
\usepackage{multirow}
\usepackage{booktabs}
\usepackage{spconf,amsmath,graphicx,hyperref}
\usepackage{xcolor}
\usepackage{pgfplots}
\usepackage{tikz}
\usepackage{booktabs}
\usepackage{tabularx}
\usepackage{placeins}
\usepackage{amssymb}
\usepackage{textcomp}
\usepackage{array, makecell}

\title{ProKWS: Personalized Keyword Spotting via Collaborative Learning of Phonemes and Prosody}
\name{
Jianan Pan, Yuanming Zhang, Kejie Huang$^*$\thanks{$^*$Corresponding author.}
}
\address{
College of Information Science and Electronic Engineering, Zhejiang University, Hangzhou, China \\
}
\begin{document}
% \ninept
%
\maketitle
\begin{abstract}
Current keyword spotting systems primarily use phoneme-level matching to distinguish confusable words but ignore user-specific pronunciation traits like prosody (intonation, stress, rhythm). This paper presents ProKWS, a novel framework integrating fine-grained phoneme learning with personalized prosody modeling. We design a dual-stream encoder where one stream derives robust phonemic representations through contrastive learning, while the other extracts speaker-specific prosodic patterns. A collaborative fusion module dynamically combines phonemic and prosodic information, enhancing adaptability across acoustic environments. Experiments show ProKWS delivers highly competitive performance, comparable to state-of-the-art models on standard benchmarks and demonstrates strong robustness for personalized keywords with tone and intent variations.
\end{abstract}
\section{Introduction}
\label{sec:intro}

Keyword Spotting (KWS) serves as a critical entry point for human–computer interaction in voice-enabled devices. Conventional systems rely on a limited set of predefined wake words (e.g., ``OK Google''), which lack flexibility and personalization. This limitation has motivated a shift toward User-Defined Keyword Spotting (UDKWS), allowing users to customize triggers without extensive data collection or model retraining, thereby enhancing user experience.

Recent advances in UDKWS have primarily focused on mitigating phonetic ambiguity, particularly in distinguishing target keywords from phonetically similar alternatives. To this end, two main paradigms have been explored: Query-by-Text (QbyT) \cite{QbyE-T1, QbyE-T2-libri-cmcd, QbyE-T3-apple, QbyE-T4-phonmatch} and Query-by-Example (QbyE) \cite{QbyE1, QbyE2, QbyE3,QbyE4, QbyE5, QbyE6}. Under the QbyT paradigm, significant progress has been achieved by learning fine-grained alignments between textual and acoustic representations. For instance, Phoneme-Level Contrastive Learning (PLCL) \cite{baseline-plcl} demonstrates that enforcing feature separation at the phoneme level enables models to achieve strong robustness against phonetically confusable phrases. Moreover, MM-KWS~\cite{mmkws} leverages multi-modal enrollments with both text and speech templates to construct more reliable keyword representations. Collectively, these efforts highlight that fine-grained, often phoneme-level, feature learning is central to state-of-the-art performance.

Despite the focus on phonetic precision, existing UDKWS systems remain both speaker- and intent-agnostic. They capture only \emph{what} is said, while neglecting \emph{how} it is spoken. Prosody (rhythm, stress, and intonation) conveys crucial information, distinguishing commands from questions and reflecting vocal style or emotional state. Without modeling prosody, UDKWS systems risk becoming unreliable, prone to misinterpreting intent, and less robust to natural variations such as accent or emotional tone. Although prosodic cues are extensively studied in speaker verification and emotion recognition, their potential for personalized and intent-aware keyword spotting remains underexplored.

To address this gap, we propose ProKWS, a dual-stream framework that integrates fine-grained phonetic analysis with personalized prosodic modeling. The architecture comprises a Phoneme Stream that learns speaker-invariant phonetic representations via contrastive learning, enhancing robustness to confusable words and a Prosody Stream that derives a compact \emph{prosodic signature} from a few enrollment samples, capturing individual intonation and rhythm. A Collaborative Fusion Module integrates the two streams, enabling keyword detection that reflects both phonetic accuracy and vocal style conformity. Our main contributions are summarized as follows: (1) We propose ProKWS, a dual-stream encoder that jointly models phonetic content and prosodic style for keyword spotting; (2) We introduce a prosodic signature mechanism that effectively captures vocal patterns and intent variations with minimal supervision; (3) We construct the \emph{Accent-KWS} and \emph{Intent-KWS} datasets for evaluating prosody-aware keyword spotting, serving as dedicated benchmarks in our study.

\begin{figure*}[hbtp]
    \centering
    \includegraphics[page=1, width=\textwidth]{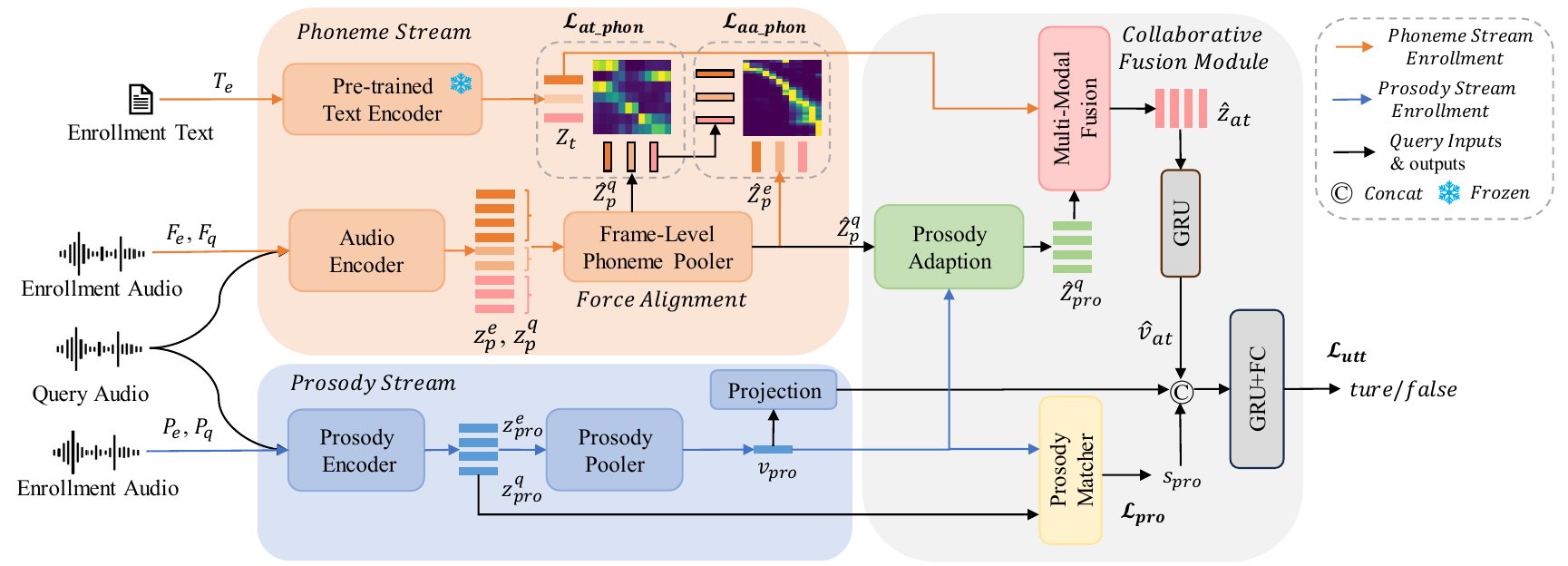}
    \caption{Overall architecture of the proposed ProKWS.}
    % The input consists of a query audio paired with either enrollment text or audio, based on the enrollment data, the output is a score used to determine whether the query matches the enrollment word.}
    \label{fig:arch}
\end{figure*}

\section{Proposed method}
\label{sec:method}
In this section, we present ProKWS, a framework for user-defined keyword spotting that integrates phonetic and prosodic information to achieve personalized and robust performance. The model processes enrollment and query inputs through parallel streams for phonetic and prosodic encoding, followed by Collaborative Fusion Module that enable dynamic interaction between the streams, as shown in Fig.\ref{fig:arch}. This design allows the system to adapt to user-specific speaking styles while maintaining fine-grained phonetic discrimination.

\subsection{Dual-Stream Encoder}
% As shown in figure1, the ProKWS architecture consists of four main components: a Phoneme Stream, a Prosody Stream, a Personalized Adaptation Module, and a Collaborative Fusion Module. 

The Phoneme Stream is designed to capture the phonetic content of speech, independent of speaker-specific characteristics. Let the batch size be \(B\). The inputs include enrollment text \(T_{\text{e}}\), along with enrollment and query audio \((X_{\text{e}}, X_{\text{q}})\). We first extract Mel-frequency Filter Bank (FBank) features \(F \in \mathbf{R}^{B \times T \times D_{\text{fbank}}}\), where \(T\) denotes the number of frames and \(D_{\text{fbank}} = 80\). These features are processed by an audio encoder, consisting of convolutional subsampling layers followed by a stack of Conformer blocks~\cite{conformer}, producing acoustic representations \(Z_{p} \in \mathbf{R}^{B \times T \times D_{p}}\), where \(D_{p}\) is the phoneme feature dimension. The same encoder is applied to both enrollment and query audio to ensure a shared feature space. Next, a pre-trained Grapheme-to-Phoneme (G2P) text encoder converts \(T_{\text{e}}\) into text embeddings \(E_{t} \in \mathbf{R}^{B \times T' \times D_{p}}\). Finally, the Phoneme Pooler aligns frame-level features with phoneme segment features \(\hat{Z}_{p}^{q} \in \mathbf{R}^{B \times T' \times D_{p}}\), using alignment information extracted by the Montreal Forced Aligner (MFA)~\cite{mfa}, which provides the start and end timestamps of each phoneme segment.

The Prosody Stream operates in parallel to extract a user’s personalized vocal style from enrollment audio. This style is represented as a fixed-size vector, termed the \emph{prosodic signature}. The input to this stream is a sequence of prosodic features, $ P \in \mathbf{R}^{B \times T \times D_{\text{pro\_in}}} $, where $ D_{\text{pro\_in}} = 3 $, corresponding to frame-level fundamental frequency (F0), aperiodicity (AP), and RMS energy. To capture temporal dependencies in these prosodic contours, the input is processed by a two-layer bidirectional GRU, referred to as the Prosody Encoder, producing $Z_{\text{pro}} \in \mathbf{R}^{B \times T \times D_{\text{pro}}} $ with $ D_{\text{pro}} = 64 $. For enrollment audio, an attention-based pooling layer, termed the Prosody Pooler, aggregates the encoder outputs \(Z_{\text{pro}}^{\text{e}}\) into a single context vector. The final output is the personalized prosodic signature $ v_{\text{pro}} \in \mathbf{R}^{B \times D_{\text{pro}}} $.

\subsection{Collabrative Fusion Module}
To incorporate a user’s speaking style into phoneme extraction, we employ a Feature-wise Linear Modulation (FiLM) layer, referred to as the Prosody Adaptation Module, to modulate the outputs of the Phoneme Stream. The personalized prosody vector $v_{\text{pro}}$ is projected into scaling and shifting vectors:
\begin{equation}
\gamma = \text{Linear}_\gamma(v_{\text{pro}}), \quad \beta = \text{Linear}_\beta(v_{\text{pro}}),
\end{equation}
yielding prosody-aware phoneme representations:
\begin{equation}
\hat{Z}_{\text{pro}}^q = \gamma \odot \hat{Z}_p^q + \beta,
\end{equation}
where the symbol $\odot$ denotes \textbf{element-wise multiplication}. We then apply a Multi-Modal Fusion Module with cross-attention to integrate the modulated phoneme features $\hat{Z}_{\text{pro}}^q$ and the text embeddings $Z_t$, where $\hat{Z}_{\text{pro}}^q$ serves as the query and $Z_t$ as both key and value:
\begin{equation}
\hat{Z}_{\text{at}} = \text{Cross-Attention}(\hat{Z}_{\text{pro}}^q, Z_t, Z_t).
\end{equation}

The Prosody Matcher applies global average pooling to the query prosody features $Z_{\text{pro}}^q$, obtaining $v_{\text{pro}}^q$, and computes the cosine similarity between $v_{\text{pro}}^q$ and $v_{\text{pro}}$, denoted as $s_{\text{pro}}$. Finally, the multi-modal fusion output vector $\hat{v}_{\text{at}}$, the personalized prosody vector $v_{\text{pro}}$, and the prosody similarity $s_{\text{pro}}$ are concatenated and passed through a GRU followed by a fully connected (FC) layer to produce the utterance-level decision score $s \in [0,1]$.

\subsection{Training Criterion}
We employ a composite objective $\mathcal{L}_{\text{total}}$:
\begin{equation}
\mathcal{L}_{\text{total}} = \mathcal{L}_{\text{utt}} + \mathcal{L}_{\text{at\_phon}} + \mathcal{L}_{\text{aa\_phon}} + \lambda \mathcal{L}_{\text{pro}},
\end{equation}
where $\lambda$ is a loss weighting hyperparameter. The utterance loss $\mathcal{L}_{\text{utt}}$ is defined as the binary cross-entropy (BCE) between the final matching score $s$ and the ground-truth label $y$:
\begin{equation}
\mathcal{L}_{\text{utt}} = -\frac{1}{B} \sum_{i=1}^B \left[ y_i \log(s_i) + (1 - y_i) \log(1 - s_i) \right].
\end{equation}

The audio-text phoneme InfoNCE loss~\cite{CPC, clap-loss} $\mathcal{L}_{\text{at\_phon}}$ contrasts aligned audio segments $z_j$ with the corresponding text embeddings $e_j$ (positives) against negative samples $e_k$:
\begin{equation}
\mathcal{L}_{\text{at\_phon}} = -\sum_j \log \frac{\exp(\text{sim}(z_j, e_j)/\tau)}{\sum_k \exp(\text{sim}(z_j, e_k)/\tau)},
\end{equation}
where $\text{sim}(\cdot, \cdot)$ denotes cosine similarity and $\tau$ is a temperature scaling hyperparameter. The audio-audio phoneme InfoNCE loss $\mathcal{L}_{\text{aa\_phon}}$ is defined analogously, contrasting paired audio segments from enrollment and query. The prosody similarity loss $\mathcal{L}_{\text{pro}}$ minimizes the distance between positive pairs:
\begin{equation}
\mathcal{L}_{\text{pro}} = \frac{1}{|B_{\text{pos}}|} \sum_{i \in B_{\text{pos}}} \left(1 - \text{sim}(v_{\text{pro}}^{q(i)}, v_{\text{pro}}^{(i)})\right),
\end{equation}
where $B_{\text{pos}}$ indexes positive samples.

\section{Experiments Configuration}

\begin{table}[t]
\centering
\caption{Negative Examples from LibriPhrase and WenetPhrase.}
% \resizebox{\linewidth}{!}{
\begin{tabular}{@{}lcccc@{}}
\toprule
 & \bf{Anchor} & \bf{Easy} & \bf{Hard} \\
\midrule
\multirow{3}{*}{LibriPhrase} & \multirow{3}{*}{friend} & guard & frind \\
 & & comfort & rend \\
 & & superior & trend \\
% \cmidrule(lr){2-4}
%  & \multirow{3}{*}{the river} & every morning & the giver \\
%  & & town with & the liver \\
%  & & not occurred & the rigor \\
\midrule
\multirow{3}{*}{WenetPhrase} & \multirow{3}{*}{ning2yuan4} & sha1mo4 & ting2yuan4 \\
 & & de2zhi1 & xing2yuan4 \\
 & & gong1wu4 & qing2yuan4 \\
% \cmidrule(lr){2-4}
%  & \multirow{3}{*}{qing1nian2ren2} & zhou1qi1xing4 & qing1nian2 \\
%  & & qiao1men2 & zhong1nian2ren2 \\
%  & & hun1ying1fa3 & qing1nian2tuan2 \\
\bottomrule
\end{tabular}
% }
\label{tab:dataset}
\end{table}

\subsection{Experimental Setups}
\subsubsection{Evaluation Datasets and Metrics}
% We evaluate our ProKWS with Google Speech Commands v1 ($\textbf{\text{G}}$)~\cite{gsc}, LibriPhrase-easy ($\textbf{\text{LP}}_\textbf{\text{E}}$) and LibriPhrase-hard ($\textbf{\text{LP}}_\textbf{\text{H}}$)~\cite{QbyE-T2-libri} in different scenarios: first, the C-KWS task, in this context, is limited to detecting specific keywords; and second, the OV-KWS task, {\it i.e.}, users can customize any keywords, unlike the C-KWS which is a closed set task; last one, the PCOV-KWS task, designed to recognize keywords that are unique to individual users.
% Below are the details about the datasets used for training and evaluation:

% \begin{itemize}
% \setlength{\itemsep}{2pt}
We evaluate ProKWS on LibriPhrase dataset, which is constructed from LibriSpeech~\cite{librispeech} \textit{train-others-500} 
in accordance with \cite{QbyE-T2-libri-cmcd, QbyE-T4-phonmatch} and divided into two parts: LibriPhrase-easy ($\textbf{\text{LP}}_\textbf{\text{E}}$) and LibriPhrase-hard ($\textbf{\text{LP}}_\textbf{\text{H}}$).
We then built WenetPhrase Easy ($\textbf{\text{WP}}_\textbf{\text{E}}$) and WenetPhrase Hard ($\textbf{\text{WP}}_\textbf{\text{H}}$) subsets following \cite{mmkws}. The samples from these subsets are shown as Table~\ref{tab:dataset}. 
% WenetPhrase comprises approximately 122K training classes and 54K test classes, totaling 2.9M samples. Each audio segment ranges from 0.5 to 2 seconds in duration and contains 2 to 6 words. 
% Additionally, we utilize Google Speech Commands v1 ($\textbf{\text{GSC}}_\textbf{\text{v1}}$) to evaluate the model on multi-classification tasks. Following the configuration in \cite{mmkws, lee2023fully}, we select 10 target keywords and the other 20 keywords are treated as unknown class, then we assess using Acc(close) and Acc(open), where Acc(close) excludes the unknown category.

For prosody-focused evaluation, we construct two TTS-generated test sets: \emph{Accent-KWS} and \emph{Intent-KWS}. Using CosyVoice2~\cite{du2024cosyvoice} for synthesis, \emph{Accent-KWS} comprises 100 hard keywords from LibriPhrase, synthesized in American, British, Indian, and Australian accents (2,400 samples total, with 3 speakers per accent and 2 utterances each). \emph{Intent-KWS} uses the same keywords but varies intents (command, question, neutral), yielding 900 samples (3 intents, 3 utterances each).

% \subsubsection{Evaluation Metrics}
We employ the Equal Error Rate (EER), where False Alarm Rate (FAR) equals False Reject Rate (FRR), and the Area Under the Curve (AUC) as critical metrics for evaluating the performance of KWS models.

% \subsubsection{Evaluation metric}
% We employ the \textbf{EER} (Equal Error Rate), where \textbf{FAR} (False Alarm Rate) equals \textbf{FRR} (False Reject Rate), and the \textbf{AUC} (Area Under the Curve) as critical metrics for evaluating the performance of KWS models. 
% Additionally, we utilize Top-1 test accuracy for assessing the performance of C-KWS systems.

\subsubsection{ Implementation details}
In our experiments, we use the AdamW optimizer~\cite{AdamW} with a weight decay of 1e-3 and an initial learning rate of 3e-4, while keeping the other params as default. The learning rate was managed by a Linear Warmup Cosine Annealing scheduler~\cite{warmup}, which linearly increased the learning rate from 0 to 3e-4 over the first 5 epochs and then decayed it using a cosine schedule~\cite{cosine_schedule} for the rest epochs.

\section{Results and Anlysis}
\label{sec:typestyle}

\begin{table}[t]
\caption{Experimental results of the proposed ProKWS on Libriphrase dataset compared to the baseline models.}
\resizebox{\linewidth}{!}{
\begin{tabular}{@{}lcccccc@{}}
\toprule
\multicolumn{1}{l}{\multirow{2}{*}{Method}} & \multirow{2}{*}{\# Params} & \multicolumn{2}{c}{AUC(\%) ↑} & \multicolumn{2}{c}{EER(\%) ↓} \\ \cmidrule(lr){3-6}
\multicolumn{1}{c}{} &                                & $\textbf{\text{LP}}_\textbf{\text{H}}$       & $\textbf{\text{LP}}_\textbf{\text{E}}$       & $\textbf{\text{LP}}_\textbf{\text{H}}$        & $\textbf{\text{LP}}_\textbf{\text{E}}$        \\ \midrule
Whisper-Tiny \cite{baseline_whisper}       & 39M      & 73.37           & 89.19             & 33.04          & 17.31     \\
Whisper-Small \cite{baseline_whisper}      & 224M     & 82.90           & 95.92             & 21.45          & 8.14      \\
Whisper-Large \cite{baseline_whisper}      & 1550M    & 85.80           & 97.54             & 19.57          & 5.33      \\ \midrule
% Triplet \cite{baseline_triplet}            & N/A      & 54.88           & 63.53             & 44.36          & 32.75     \\
CMCD \cite{QbyE-T2-libri-cmcd}                  & 0.7M     & 73.58           & 96.70             & 32.90          & 8.42      \\
CLAD \cite{baseline-clcd}                  & 2.2M     & 76.15           & 97.03             & 30.30          & 8.65      \\ 
EMKWS \cite{baseline_emkws}                & 3.7M     & 84.21           & 97.83             &  23.36          & 7.36      \\
PhonMatchNet \cite{QbyE-T4-phonmatch}  & 0.7M     & 88.52           & 99.29             & 18.82          & 2.80      \\
CED \cite{baseline_ced}                    & 3.6M     & 92.70           & 99.84             & 14.40          & 1.70      \\
AdaKWS-Tiny \cite{baseline_adakws}         & 15M      & 93.75           & 99.80             & 13.47          & 1.61      \\
% AdaKWS-Small \cite{baseline_adakws}        & 109M     & 95.09           & 99.82             & 11.48          & 1.21      \\
MM-KWS \cite{mmkws}                          & 3.9M     & 96.25  & 99.95    & 9.30  & 0.68      \\ 
PLCL  \cite{baseline-plcl}                              & 3.9M     & 96.59  & \textbf{99.97}    & 8.47  & \textbf{0.57}      \\ \midrule
ProKWS                                   & 2.9M     & \textbf{96.92}  & 99.96    & \textbf{7.52}  & 0.63      \\ \bottomrule
\end{tabular}
}
\label{tab:LibriPhrase}
\end{table}

\subsection{Comparative Evaluation of ProKWS}

The experimental results of the proposed ProKWS model on both the standard benchmark and synthesized datasets, compared with baseline models, are presented in Table~\ref{tab:LibriPhrase} and Table~\ref{tab:tts}. 
% The results show that ProKWS delivers highly competitive performance, comparable to state-of-the-art systems on standard benchmarks, while also demonstrating strong robustness for personalized keywords under tone and intent variations. 
% Specifically, ProKWS achieves an AUC of 99.96\% and an EER of 0.63\% on the $\textbf{LP}_\textbf{E}$ dataset, and an AUC of 96.92\% with an EER of 7.52\% on the $\textbf{LP}_\textbf{H}$ dataset.

\begin{table}[t]
\caption{Experimental results of ProKWS on the WenetPhrase, \emph{Accent-KWS} (AC) and \emph{Intent-KWS} (IT) dataset compared to the baseline (BL).}
\resizebox{\linewidth}{!}{
\begin{tabular}{@{}lcccccccc@{}}
\toprule
\multirow{2}{*}{} & \multicolumn{4}{c}{AUC(\%) ↑} & \multicolumn{4}{c}{EER(\%) ↓} \\ 
\cmidrule(lr){2-9}
& $\text{WP}_\text{H}$ & $\text{WP}_\text{E}$ & AC & IT & $\text{WP}_\text{H}$ & $\text{WP}_\text{E}$ & AC & IT \\ 
\midrule
BL \cite{mmkws} & \textbf{85.84} & 99.15 & 52.39 & 61.35 & \textbf{22.06} & 4.26 & 47.23 & 37.67 \\
Ours & 84.82 & \textbf{99.81} & \textbf{71.45} & \textbf{86.42} & 23.33 & \textbf{1.84} & \textbf{27.92} & \textbf{18.10} \\ \bottomrule
\end{tabular}
}
\label{tab:tts}
\end{table}

\begin{figure}[t]
    \centering
    \includegraphics[width=\columnwidth]{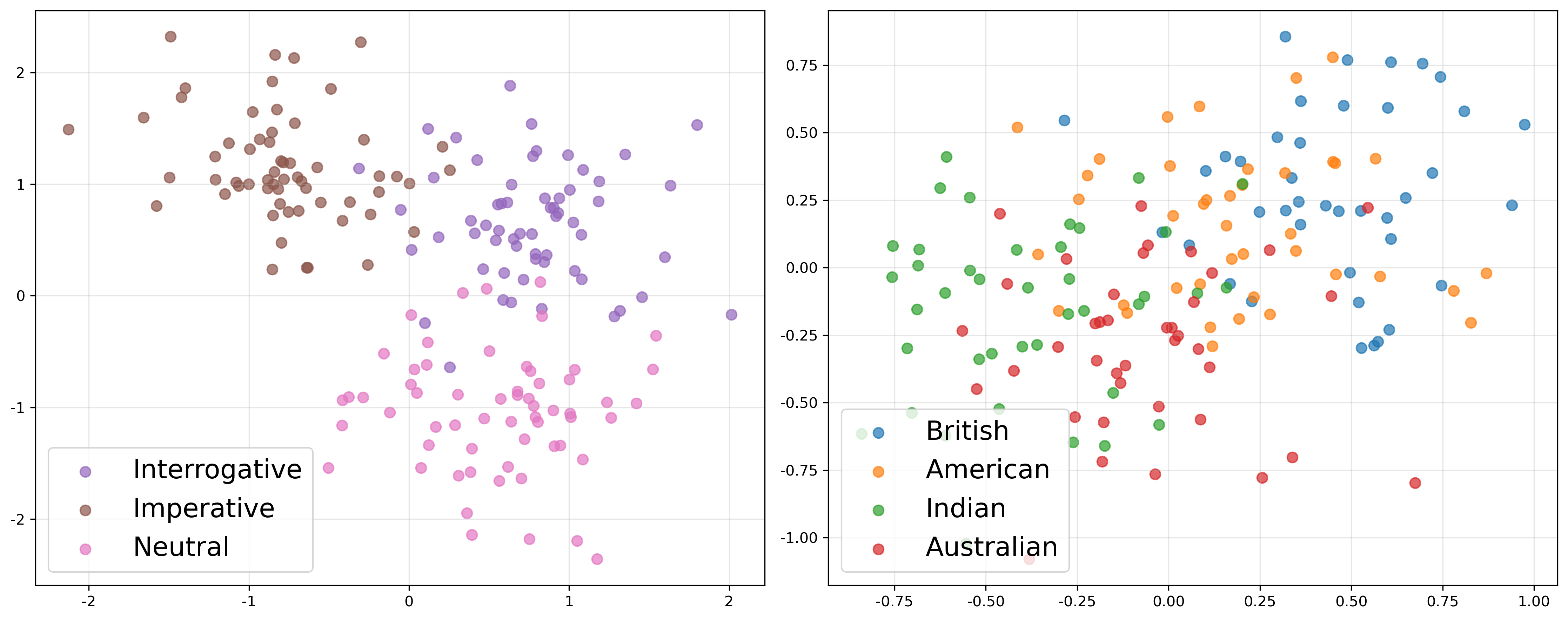}
    \caption{t-SNE visualization of prosodic signatures across different accents and intents.}
    \label{fig:clusters}
\end{figure}

\begin{figure}[t]
    \centering
    \includegraphics[width=\columnwidth]{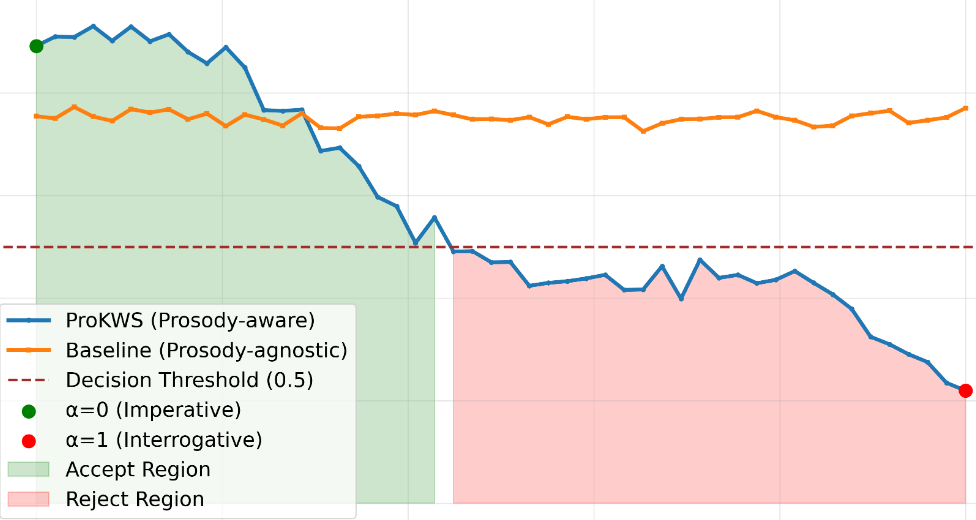}
    \caption{Score variation analysis for continuous intent change. The x-axis represents the interpolation coefficient $\alpha$ between imperative and interrogative prosody, and the y-axis represents the resulting score $s(\alpha)$.}
    \label{fig:boundary}
\end{figure}

\subsection{Visual Analysis}

To examine what the proposed Prosody Stream has learned and how it contributes to the final decision, we first use t-SNE to project the learned prosodic signature vectors $v_{pro}$. As shown in Fig.~\ref{fig:clusters}, for the same keyword in the Intent-KWS dataset, embeddings form three distinct and well-separated clusters corresponding to imperative, interrogative, and neutral intents. In contrast, embeddings exhibit weaker discrimination for accent, showing significant overlap and confusion. We attribute this to two factors: first, the prosodic features may not fully capture the stable rhythmic and intonational patterns necessary for accent characterization; second, CosyVoice exhibits limited stability in consistently generating audio with distinct accent variations.

Next, we enroll an imperative version of ``Turn on light'' and select two query audios: a positive query (imperative ``Turn on light'') and a hard-negative query (interrogative ``Turn on light''), extracting their prosody vectors $v_{pos}$ and $v_{neg}$. We then generate a series of interpolated prosody vectors:
\begin{equation}
v_{\text{interp}}(\alpha) = (1 - \alpha) v_{\text{pos}} + \alpha v_{\text{neg}} \quad \text{for} \quad \alpha \in [0,1].
\end{equation}
Each interpolated vector $v_{\text{interp}}(\alpha)$ is fed into the frozen Fusion Module to compute the final score $s(\alpha)$. As illustrated in Fig.~\ref{fig:boundary}, the score produced by ProKWS remains high when the prosody closely matches the enrollment sample, but decreases as it shifts toward the mismatched intent. In contrast, a baseline model without a prosody stream maintains a consistently high score across the interpolation, indicating insensitivity to prosodic variations. This analysis visually demonstrates that the Prosody Stream enables the model to better distinguish fine-grained intent variations.

\subsection{Ablation Studies of ProKWS}

We conduct ablation studies to evaluate the contribution of each component in ProKWS, as shown in Table~\ref{tab:ablation}. Removing the Prosody Adaptation Module significantly increases EER on $\textbf{\text{LP}}_\textbf{\text{H}}$ ($7.52\% \rightarrow 12.29\%$), highlighting its importance in adaptively fusing prosody with phoneme features. Excluding $\mathcal{L}_{\text{pro}}$ leads to a further performance drop, confirming its role in learning speaker-invariant prosody. Eliminating the entire prosody stream causes the largest degradation, demonstrating that prosodic cues are critical for capturing both intent and speaker-specific variations.

\begin{table}[t]
\caption{Ablation studies of ProKWS}
\begin{center}
\resizebox{\linewidth}{!}{
\begin{tabular}{ccccc}

\hline
\multirow{2}{*}{\textbf{Method}} &  \multicolumn{2}{c}{\textbf{AUC(\%)}$\uparrow$}             & \multicolumn{2}{c}{\textbf{EER(\%)}$\downarrow$}            \\ \cline{2-5} 
      & $\mathrm{\textbf{LP}}_\mathrm{H}$ & $\textbf{LP}_\textbf{E}$ & $\textbf{LP}_\textbf{H}$ & $\textbf{LP}_\textbf{E}$  \\ \hline
\multicolumn{1}{l}{ProKWS} &\textbf{96.92}&\textbf{99.96}  &\textbf{7.52}&\textbf{0.63} \\
\multicolumn{1}{l}{\hspace{2em}w/o Prosody Adaption Module} &      94.22&99.67&12.29&1.86   \\
\multicolumn{1}{l}{\hspace{4em}w/o auxiliary $\mathcal{L}_{\text{pro}}$} &      92.76&99.34&13.47&3.24   \\
\multicolumn{1}{l}{\hspace{6em}w/o Prosody Stream} & 
88.71&98.99&15.34&4.67   \\
\hline
\end{tabular}
}
\label{tab:ablation}
\end{center}
\end{table}

\section{Conclusion}
This paper presented ProKWS, a dual-stream framework that jointly models phonemes and prosody for personalized keyword spotting. By integrating phoneme-level contrastive learning with prosodic signatures, ProKWS achieves strong results on standard benchmark datasets and demonstrates superior robustness on our synthesized \emph{Accent-KWS} and \emph{Intent-KWS} benchmarks. Visualization analyses confirm the discriminative power of prosodic embeddings, while ablation studies validate the necessity of each component. Overall, ProKWS advances user-defined keyword spotting by unifying phonetic accuracy with prosodic personalization, paving the way for intent-aware and user-adaptive KWS systems.

%\vfill
%\pagebreak

% \section{COPYRIGHT FORMS}
% \label{sec:copyright}

% You must submit your fully completed, signed IEEE electronic copyright release
% form when you submit your paper. We {\bf must} have this form before your paper
% can be published in the proceedings.

\vfill\pagebreak
\ninept
\section{Acknowledgements}
% \label{sec:refs}
This work was supported by the National Science and Technology Major Project (2026ZD1305800), Zhejiang Province's Leading Talent Project in Science and Technology Innovation (2023R5204) and Ningbo Key Research and Development Program (2025Z082).

% -------------------------------------------------------------------------
\bibliographystyle{IEEEbib}
\bibliography{strings,refs}

\end{document}